\documentclass[twocolumn,preprintnumbers,aps,prd,superscriptaddress,nofootinbib,tightenlines]{revtex4-1}

\usepackage{amsmath, amsfonts, amsthm, amssymb, graphicx}
\usepackage{xcolor}
\usepackage[utf8]{inputenc}
\usepackage[normalem]{ulem} %%%%%This is just required for the strike-through change marking and can be removed for final version!
\def\be{\begin{equation}}
\def\ee{\end{equation}}
\def\ba{\begin{eqnarray}}
\def\ea{\end{eqnarray}}

\def\bq{\begin{quote}}
\def\eq{\end{quote}}

 at 10truept

\newcommand{\beq}{\begin{equation}}
\newcommand{\eeq}{\end{equation}}
\newcommand{\bea}{\begin{eqnarray}}
\newcommand{\eea}{\end{eqnarray}}
\newcommand{\beqa}{\begin{eqnarray}}
\newcommand{\eeqa}{\end{eqnarray}}

\newcommand{\Poincare}{Poincar\'{e}~}

\begin{document}
\raggedbottom

\title{A No-go Theorem for a Gauge Vector as a Space-time Goldstone}

\author{Remko Klein} 
\email{remko.klein@rug.nl}
\affiliation{Van Swinderen Institute for Particle Physics and Gravity, University of Groningen, \\ Nijenborgh 4, 9747 AG Groningen, The Netherlands} 
\author{Emanuel Malek} 
\email{e.malek@lmu.de}
\affiliation{Arnold Sommerfeld Center for Theoretical Physics, Department f\"ur Physik, \\ Ludwig-Maximilians-Universit\"at M\"unchen, Theresienstra{\ss}e 37, 80333 M\"unchen, Germany} 
\author{Diederik Roest} 
\email{d.roest@rug.nl}
\affiliation{Van Swinderen Institute for Particle Physics and Gravity, University of Groningen, \\ Nijenborgh 4, 9747 AG Groningen, The Netherlands}
\author{David Stefanyszyn} 
\email{d.stefanyszyn@rug.nl}
\affiliation{Van Swinderen Institute for Particle Physics and Gravity, University of Groningen, \\ Nijenborgh 4, 9747 AG Groningen, The Netherlands}

\preprint{CERN-TH-2018-143}
\preprint{LMU-ASC 30/18}

\date{\today}

\begin{abstract}
\noindent
Scalars and fermions can arise as Goldstone modes of non-linearly realised extensions of the Poincar\'{e} group (with  important implications for the soft limits of such theories): the Dirac-Born-Infeld scalar realises a higher-dimensional Poincar\'{e} symmetry, while the Volkov-Akulov fermion corresponds to super-Poincar\'{e}. In this paper we classify extensions of the Poincar\'{e} group which give rise to a vector Goldstone mode instead. Our main result is that there are no healthy (ghost free) interacting $U(1)$ gauge theories that non-linearly realise space-time symmetries beyond gauge transformations. This implies that the structure of e.g.~Born-Infeld theory is not fixed by symmetry.
\end{abstract}

\maketitle

\section{Introduction}

Non-linear realisations of spontaneously broken symmetries form an important and interesting part of quantum field theory. When global internal symmetries are broken, Goldstone's theorem tells us that there is a massless mode for every broken generator \cite{Goldstone1, Goldstone2} (with an adapted counting in non-relativistic systems, see e.g. \cite{Watanabe1,Watanabe2}). The non-linear transformation rules and invariants can be efficiently extracted from the coset construction \cite{internal1,internal2}. 

In contrast, Goldstone's theorem does not apply to spontaneously broken space-time symmetries \cite{spacetime1,spacetime2}, for which there can be fewer Goldstone modes than broken generators: every spontanously broken generator that commutes with translations into another such generator gives rise to an \textit{inessential} Goldstone  \cite{redundant1,redundant2}. The latter can be removed from the low-energy effective field theory (EFT)  by imposing inverse Higgs constraints \cite{spacetime2}. Alternatively, one can integrate them out of the path integral since the inessential modes acquire a mass gap. In many cases, and possibly all, these two possibilities lead to equivalent EFTs for the \textit{essential} Goldstones \cite{KRS} which non-linearly realise the full broken symmetry group.

The impact of these non-linearly realised symmetries on physical observables is beautifully captured by soft limits of scattering amplitudes. A very concrete and simple example is Adler's zero \cite{adler,adler2}: scattering amplitudes involving Goldstone modes of internal symmetries vanish in the limit where a single external momentum is taken soft, i.e.~the Taylor expanded amplitude begins at linear order in the soft momenta.

In the case of a single Goldstone scalar, there are specific EFTs that display a further enhancement of the soft scaling to quadratic or cubic order \cite{softamp1,softamp2}. The quadratic scaling can be traced to a non-linearly realised Poincar\'{e} group in one higher dimension, or its contraction dubbed the Galileon group \cite{wess}; i.e. to space-time symmetry groups. The cubic scaling involves a further extension of the Galileon group \cite{specialgalgroup1,specialgalgroup2}, and a soft scaling beyond cubic order is not possible. The above list of enhanced scalings and space-time symmetries can be proven to be exhaustive from the soft limit \cite{softamp1,softamp2,softamp3} and Lie-algebraic \cite{Brauner} perspective.

A similar analysis has been initiated in the context of fermionic Goldstones, or ``Goldstinos''. The known examples correspond to the Volkov-Akulov (VA) fermion \cite{VA}, which non-linearly realises $\mathcal{N} = 1$ supersymmetry, and a shift symmetric fermion which non-linearly realises a contraction of the supersymmetric algebra \cite{Fshift}. The soft amplitudes for these theories also exhibit special behaviour \cite{VAamps1,VAamps2,brando}. 

For both the scalar and fermion, one therefore has well-defined space-time symmetry breaking patterns that are in one-to-one correspondence with special soft behaviour. It is natural to wonder about the extension to vector modes. In this paper we will provide a Lie-algebraic study regarding possible non-linearly realised space-time symmetries for a vector Goldstone. We will comment on complementary amplitude results in the discussion.

The outline of this paper is as follows. We classify different space-time algebras that give rise to a vector Goldstone in section II. The implications for effective field theories and the role of gauge symmetry is discussed in section III. There we show that a healthy (ghost free) self-interacting $U(1)$ gauge vector cannot be a space-time Goldstone with non-linear symmetries that go beyond gauge transformations. We present our conclusions and outlook in section IV.

\section{Lie-Algebra Classification}

We are interested in $D$-dimensional relativistic field theories of a single vector Goldstone, and will therefore construct groups $G$ which include a linearly realised Poincar\'{e} subgroup  with generators $P_{\mu}$ and $M_{\mu\nu}$ as well as a non-linearly realised vector generator $Q_\mu$. Additional non-linearly realised generators can be added provided their Goldstone modes are inessential due to inverse Higgs constraints that set (a projection of) a $G$-covariant derivative to zero. 

It will be useful to introduce some terminology here. We label the essential Goldstone as $G_{0}$, while inessentials that are solved for using the essential Goldstone's covariant derivative we refer to as first-order inessentials $G_1$. We inductively define an $n^{\text{th}}$ order inessential $G_n$ as one which is eliminated by the covariant derivative of an $(n-1)^{\text{th}}$ order inessential. This assumption amounts to commutators between translations and $G_{n}$ taking the form  (see \cite{spacetime2,KRS} for more details) 
\begin{equation}
[P,G_{n}] = G_{n-1} + \text{Poincar\'{e}}. \label{higher-order-inessentials}
\end{equation}
We exclude algebras where all inessentials cannot be uniquely assigned an order in this way. To our knowledge no such algebras have been constructed.

\bigskip

Let us first see what the ordering \eqref{higher-order-inessentials} implies for the subset $P_{\mu}, M_{\mu\nu}$ and $Q_{\mu}$. In the absence of Levi-Civita tensors (more on which in the concluding section), Lorentz invariance fixes the following form for the only non-trivial commutators (all others are specified by having a Poincar\'{e} factor and by $Q_\mu$ transforming as a Lorentz vector):
 \begin{align}
[ P_\mu , Q_\nu ] = a M_{\mu \nu}  \,, \quad    [ Q_\mu , Q_\nu ] = b M_{\mu \nu} + \sum_{i=1} c_{i} N_{\mu\nu}^{(i)}  \,,
  \end{align}
where $a,b,c_{i}$ are real constants and the two-forms $N_{\mu\nu}^{(i)}$ can correspond to Lorentz projections of higher-order generators with more indices. Jacobi identities imply $a=b=c_{i} = 0$, ensuring that $P_{\mu}, M_{\mu\nu}$ and $Q_{\mu}$ always form a subalgebra even when we add a number of higher-order generators. Throughout this paper we will therefore have the doublet of generators  
 \begin{align}
 \vec{U}_{\mu} = (P_\mu, Q_\mu)^{\rm T} \,, \label{generator-doublet}
 \end{align}
 commuting amongst itself.

The non-linear transformation rules for the Goldstones are extracted by left multiplication of the coset element by an element of the full symmetry group. Without inessentials, the coset element is\footnote{We will assume that the proof of coset universality  for internal symmetries also applies to space-time symmetries.} 
\begin{equation}
\gamma = e^{x^{\mu}P_{\mu}}e^{A^{\mu}Q_{\mu}} \,,
\end{equation}
and given that
\begin{equation}
e^{q^{\mu}Q_{\mu}}\gamma = e^{x^{\mu}P_{\mu}}e^{(A^{\mu} + q^{\mu})Q_{\mu}} \,,
\end{equation}
the broken generator $Q_{\mu}$ induces a constant shift on the vector $\delta A_\mu = q_\mu$. 

This result is the first marked difference compared to the scalar or fermion Goldstone case, in which it is possible to have a non-trivial transformation beyond a constant shift in the absence of higher-order generators. In the scalar case we have a scalar generator, $X$, with $[P_{\mu},X]$ distinguishing between two possibilities. The commutator can be proportional to $P_\mu$, in which case the essential Goldstone is the dilaton and generates a space-time symmetry. Instead, taking the commutator to vanish implies that $X$ induces a shift symmetry on the essential Goldstone and corresponds to an internal symmetry. 

For a fermion, a similar role is played by the anti-commutator $\{Q, \bar{Q} \}$ between the fermionic generators. This can be proportional to a translation, leading to supersymmetry; i.e. supersymmetry transformations are the `square-root' of translations. The anti-commutator can also vanish, in which case supersymmetry is contracted to a shift symmetry for a fermion \cite{Fshift}. The above vector result implies that one cannot take the `square-root' of Lorentz transformations in a similar fashion\footnote{The situation in $D=3$ is different: it allows for a non-trivial vector transformation involving a Levi-Civita tensor \cite{Bansal}.}.

\bigskip

In order to have non-trivial vector transformations beyond the constant shift, we have to include inessential Goldstones. The extension of the previous subalgebra with first-order generators again forms a subalgebra, consisting of \Poincare as well as $G_0$ and $G_1$. This follows straightforwardly from Jacobi identities. Consider the commutator $[ G_0, G_1 ]$ and the associated Jacobi identity with $P_\mu$ which implies
\begin{equation}
[P_{\mu},[G_0, G_1]] = 0.
\end{equation}
Given that any higher-order generators must have a non-vanishing commutator with translations, we infer that only the doublet $P_\mu$ and $Q_\mu$ can appear on the right hand side of $[G_0, G_1]$. A similar argument can be applied to the remaining relevant commutators to complete the subalgebra proof. As we will discuss at the end of this section, this fact plays an important role in our no-go.

First-order inessentials, by definition, can be eliminated by setting a projection of the $G$-covariant derivative of the vector to zero. Since the latter has three possible Lorentz projections, we can extend our algebra with one anti-symmetric two-form $N_{\mu\nu}$, one symmetric and traceless tensor $S_{\mu\nu}$, one scalar $T$ or a combination of these three generators. The most general algebra with these generators has the following form. In addition to the commuting doublet, the crucial commutators are between the first-order generators and the vector doublet (where (anti-)symmetrisation is with weight $1$)
 \begin{align}
  [\vec{U}_{\mu},N_{\rho \sigma}] &= 2\mathcal{M}_N \eta_{\mu [\sigma}\vec{U}_{\rho]} \,, \notag \displaybreak[1] \\
   [\vec{U}_{\mu},S_{\rho \sigma}] &= \mathcal{M}_S \left(2 \eta_{\mu (\sigma}\vec{U}_{\rho)} - \frac{2}{D}\eta_{\rho \sigma} \vec{U}_{\mu} \right) \,, \notag \\
  [\vec{U}_{\mu}, T] & = \mathcal{M}_T \vec{U}_{\mu} \,,
 \end{align}
where $\mathcal{M}_N$ is a $2$x$2$ matrix in doublet space with entries $\nu_{i=1, \ldots, 4}$ and $\mathcal{M}_S$ and $\mathcal{M}_T$ are defined similarly in terms of $\sigma_i$ and $\tau_i$, respectively. The ordering condition \eqref{higher-order-inessentials} implies that the second of these entries (in the upper right position) has to be non-vanishing in order for the corresponding Goldstone to be inessential. Without loss of generality it will be set to unity. 

The structure of the algebra and its non-linear realisation is determined by the properties of the matrices ${\cal M}_{N,T,S}$. To extract the action of the broken generators we parametrise the coset element as
\begin{equation}
\gamma = e^{x^{\mu}P_{\mu}}e^{A^{\mu}Q_{\mu}}e^{B^{\mu \nu} N_{\mu \nu} + g^{\mu \nu} S_{\mu \nu} + \phi T} \,.
\end{equation}
Note that our coset parametrisation has the generators of each order appearing in a separate exponential with the higher-order inessentials to the right. We will denote the coset coordinates corresponding to the doublet $P_\mu$ and $Q_\mu$ by
\begin{align}
  \vec{V}^{\mu} = (x^\mu, A^\mu)^T \,.
 \end{align}
By computing the left multiplication of this coset element to leading order in the parameters $n_{\mu \nu}, s_{\mu \nu}, t$ of the broken  generators, we find the following infinitesimal transformation rules\footnote{Since these transformation rules are independent of the inessential Goldstones, the vector EFT is the same regardless of whether we eliminate the inessentials with an inverse Higgs constraint or integrate them out of the path integral.}
\begin{equation}
\delta \vec{V}^{\mu} = ( \mathcal{M}_N^{T} n^{\mu}{}_{\nu}  + \mathcal{M}_S^{T} s^{\mu}{}_{\nu}  + \mathcal{M}_T^{T} t \delta^\mu_\nu ) \vec{V}^{\nu} \,,
\end{equation}
which, as co-vectors, involve the transposed matrices. In the active form
\begin{align}
\delta A_\mu = & n_{\mu\nu} ( x^{\nu} + \nu_4 A^{\nu}) - n_{\sigma \rho}(\nu_1 x^{\rho} + \nu_3 A^{\rho})\partial^{\sigma}A_{\mu} + \notag \\
 & s_{\mu\nu} ( x^{\nu} + \sigma_4 A^{\nu}) - s_{\rho \sigma}( \sigma_1 x^{\rho} + \sigma_3 A^{\rho})\partial^{\sigma}A_{\mu} + \notag \\
& t ( x^{\mu} + \tau_4 A^{\mu} - (\tau_1 x^{\nu} + \tau_3 A^{\nu})\partial_{\nu}A^{\mu}) \,, \label{scalarcoset}
\end{align}
while the coordinates do not transform. Note that transformations with $\nu_1 = \nu_4$ are Lorentz transformations, and hence we will take $\mathcal{M}_N$ to be traceless without loss of generality. Similarly, the trace of $\mathcal{M}_T$ scales $x^\mu$ and $A^\mu$ evenly, leaves the field strength $F_{\mu \nu} = 2 \partial_{[\mu}A_{\nu]}$ invariant and will play no role in what follows.

In order to close the algebra, the commutators between the first-order generators are given by
\begin{align}
& [N,N] \sim M + N \,,  \qquad\,\,\,\, [S,S] \sim M + N \,,  \label{structures}  \\
& [N,S] \sim M +  N + S \,, \quad [N,T] \sim M + N \,, \quad [S,T] \sim S \,, \notag
\end{align}
where we have supressed Lorentz structures since all terms on the RHS correspond to a unique structure. Jacobi identities impose the following constraints on the matrices
 \begin{align}
  & \mathcal{M}_N^2 = a_1 \mathbb{I}_2 + a_2 \mathcal{M}_N \,, \quad \mathcal{M}_S^2 = b_1 \mathbb{I}_2 + b_2 \mathcal{M}_N \,, \notag \\
  & [ \mathcal{M}_N, \mathcal{M}_S ] = c_1 \mathbb{I}_2 + c_2 \mathcal{M}_S + c_3 \mathcal{M}_N \,, \notag \\
  & [\mathcal{M}_N, \mathcal{M}_T ] = d_1 \mathbb{I}_2 + d_2 \mathcal{M}_N \,, \notag \\
  & [\mathcal{M}_S, \mathcal{M}_T ] = e_1 \mathcal{M}_S \,, \label{Jacobi}
\end{align}
where the coefficients on the RHS parametrise the different terms of the RHS of \eqref{structures}. One can still perform basis changes of the doublet of generators \eqref{generator-doublet} to simplify the matrices; however, one can only do this to bring a single matrix to a preferred form.

The most general solution to the Jacobi identities for the algebra with three first-order generators has 
 \begin{itemize}
 \item
 the traceless parts of the three matrices equal and arbitrary, which can be brought to the form $(0,1; s,0)$ with $s=0,\pm1$, plus an arbitrary trace $(\lambda,0;0,\lambda)$ for $\mathcal{M_S}$, or
 \item
 the three matrices equal and with vanishing determinant, which can be brought to the form $(0,1;0,0)$, plus an arbitrary traceless diagonal $(\lambda,0;0,-\lambda)$ for either $\mathcal{M}_S = \mathcal{M}_T$, $\mathcal{M}_{N} = \mathcal{M}_T$ or $\mathcal{M}_{T}$. 
\end{itemize}
In all cases, the coefficients in \eqref{Jacobi} are determined by the matrices ${\cal M}_{N,T,S}$. 

Smaller algebras with fewer than three first-order generators can be classified similarly, by solving \eqref{Jacobi} with some matrices vanishing, with essentially the same results as above. For example, when the only first-order generator is the two-form, the first equation of \eqref{Jacobi} allows for arbitrary $\mathcal{M}_N$ and fixes $a_{1,2}$ in terms of its determinant and trace. 

Finally, the presence of second- and higher-order generators will not change the transformation rules induced by the first-order generators. This follows from the fact that the above algebras are always subalgebras and from our choice of coset parametrisation.

\section{Gauge symmetry}

We now turn to the physical theories exhibiting such symmetries. Importantly, the shift symmetry of the vector requires it to be derivatively coupled, e.g.~ forbidding a mass term $m^2 A_\mu^2$. Therefore, the longitudinal mode does not have a healthy kinetic term, and it is either infinitely strongly coupled or a (massless) ghost. Healthy vector theories with interesting non-linear symmetries will therefore have to feature gauge invariance.

We follow \cite{gauge1,gauge2,gauge3} to embedd a $U(1)$ gauge symmetry in the coset construction as follows. The gauge symmetry includes an infinite number of global symmetries of the form
\begin{equation}
\delta A_{\mu} = \sum_{n=1}^{\infty}  s_{\mu \nu_{2} \ldots \nu_{n}}x^{\nu_{2}}\ldots x^{\nu_{n}} \,, \label{gaugesymm}
\end{equation}
where the parameters $s_{\mu \nu_{2} \ldots \nu_{n}}$ are symmetric constants (which include both traceless and trace parts). These transformations can be derived from the coset construction by augmenting the Poincar\'{e} group with an infinite number of fully symmetric generators\footnote{While gauge symmetry can be implemented in the coset construction in an analogous manner to non-linear (space-time) symmetries, its implications are fundamentally different: the latter impose restrictions on the interactions of a specific degree of freedom, while the former eliminates the longitudinal mode altogether. In this terminology, a gauge vector without additional non-linear symmetries therefore does not constitute a Goldstone mode.}  $S_{\mu \nu_{2} \ldots \nu_{n}}$. For $n=1$ we have a shift symmetry and hence $S_\mu$ corresponds to our essential vector generator $Q_{\mu}$. The higher-order generators only have the non-trivial commutation relation\footnote{The above can be easily adapted for a theory of a massive vector which, in the St\"{u}ckelberg formalism, includes a longitudinal scalar with transformation rules $A_{\mu} \rightarrow A_{\mu} + \partial_{\mu}\Lambda$, $\phi \rightarrow \phi - \Lambda$. To realise this, we add a scalar generator $X$ and the $n=1$ commutator $[P_{\mu},S_{\nu}] = \eta_{\mu\nu}X$ to \eqref{gaugecomms}.} 
\begin{align}
[P_{\mu}, S_{\nu \rho_{2} \ldots \rho_{n}} ] = (n-1) \eta_{\mu (\nu} S_{\rho_{2} \ldots \rho_{n})} \label{gaugecomms} \,,
\end{align}
as required by the inverse Higgs ordering \eqref{higher-order-inessentials}.

\bigskip

The introduction of gauge symmetry implies specific transformation rules for the first-order inessential subalgebra of the previous section: it implies that now the algebra matrices $\mathcal{M}_S$ and $\mathcal{M}_T$ are given by the degenerate case $(0,1;0,0)$. The most general solution to the Jacobis therefore only allows for the remaining first-order inessential, generated by the anti-symmetric two-form $N_{\mu \nu}$, to have the same matrix, ${\cal M}_N$. Thus, the only two-form transformation compatible with gauge symmetry is given by 
 \begin{equation}
  \delta A_{\mu} = n_{\mu\nu}x^{\nu} \label{vector-galileon} \,,
 \end{equation}
and hence the field strength $F_{\mu \nu}$ shifts with the constant two-form parameter. 
One would naturally call any theory invariant under this symmetry a \textit{vector Galileon} due to its similarity with the scalar Galileon \cite{galileon}: in both cases a shift symmetry is accompanied by a shift linear in the space-time coordinates.

The Maurer-Cartan form in this case takes a very simple structure. Once we have solved for the inessential two-form Goldstone by setting the anti-symmetric part of the vector's covariant derivative to zero, all $U(1)$ gauge invariant self-interactions are constructed from $\partial_{\sigma}F_{\mu\nu}$ and its derivatives. Each of these will lead to an unhealthy theory with an Ostrogradski ghost in the spectrum. Any healthy invariant interactions must therefore be Wess-Zumino terms. However, it has been proven that no such terms exist \cite{nogo}. Given that \eqref{vector-galileon} does not change when we add second- and higher-order generators, we therefore conclude that a gauge vector has no healthy self-interactions that non-linearly realise a space-time symmetry beyond gauge transformations when we include the two-form generator. 

\bigskip

The remaining possibility for healthy interacting $U(1)$ theories with non-trivial space-time symmetries therefore consists of having gauge symmetry augmented with non-symmetric inessentials at higher-orders. However,   we will argue that the higher-order inessentials do not change the story. In particular, without the first-order two-form, the only higher-order inessentials one can add correspond to gauge transformations. For concreteness, we will concentrate on second-order inessentials but our argument holds at any order. 

Since we are omitting the first-order two-form, the second-order symmetric, traceful gauge generator $S_{\mu \nu \rho}$ can only be augmented with a hook generator $H_{\mu \nu ; \rho}$ (with $H_{\mu \nu ; \rho} = - H_{\nu \mu; \rho}$ and $H_{[\mu \nu; \rho]} = 0$); the only other option is the anti-symmetric three-form but since this does not correspond to any projections of the first-order inessentials covariant derivatives, its corresponding Goldstone cannot be inessential. However, the Jacobi identities imply that the Goldstone corresponding to the hook generator cannot be inessential either: in the absence of the two-form, Jacobi identities require $[P_{\mu},H_{\nu\rho;\sigma}] = 0$. This can be seen most easily at the level of transformation rules. If the hook Goldstone  was inessential then it would induce a transformation on the essential vector of the form
 \begin{align}
  \delta_h A_\mu = h_{\mu \nu; \rho} x^\nu x^\rho + \ldots \,,
 \end{align}
where we have omitted field-dependent terms. Upon commuting this with translations ($\delta_{\epsilon}A_{\mu} = -\epsilon^{\nu}\partial_{\nu}A_{\mu}$), we obtain
 \begin{align}
  [ \delta_\epsilon , \delta_h ] A_\mu = h_{\mu \nu;\rho} (x^\nu \epsilon^\rho + \epsilon^\nu x^\rho) + \ldots \,.
 \end{align}
The latter is however not a gauge transformation and hence would require the presence of the first-order two-form generator for the algebra to close.

This analysis can be repeated order by order to show that in the absence of the first-order two-form generator, the only higher-order generators whose corresponding Goldstones can be eliminated by an inverse Higgs constraint correspond to a $U(1)$ gauge transformation. 

\section{Discussion \& Outlook}

For scalars and fermions, the possible non-linear realisations of space-time symmetries are always accompanied by enhanced soft limits, and vice versa. In this paper we have addressed the question of whether similar symmetries are possible for vectors as well. Our main result is that none of the possible algebras with a single essential vector Goldstone (as classified in section II) that are compatible with gauge symmetry allow for healthy interacting theories (as proven in section III). We have also shown that adding higher-order inessential Goldstones does not change the story.

Throughout our derivation we have assumed coset universality as well as the inverse Higgs ordering \eqref{higher-order-inessentials} as satisfied by all known examples. We have also assumed the absence of Levi-Civita tensors in the algebra such that our results are valid in arbitrary dimensions. However, in $D=4$ we have checked that adding Levi-Civita tensors does not affect our no-go. Indeed, with Levi-Civita tensors in the commutators the vector generator still induces a shift symmetry (since we still have $[\vec{U}_{\mu},\vec{U}_{\nu}] = 0$) and at the first-order level Jacobi identities only allow for Levi-Civita dependence when the first-order inessential is a two-form but again the new terms do not allow for any healthy interactions.  

Remarkably, the question whether the structure of a gauge theory can be fixed by a non-linear symmetry was answered virtually simultaneously from a complementary amplitude perspective, with the same negative result \cite{Elvang-vectors}. Note that this also applies to the Born-Infeld (BI) theory of a gauge vector. In that case, the absence of a non-linear symmetry  follows from writing BI in terms of the metric 
 \begin{align}
   g_{\mu \nu} = \eta_{\mu \nu} + F_{\mu \nu} \,, \label{metric}
  \end{align}
where, in contrast to the induced metric of the Dirac-Born-Infeld scalar and Volkov-Akulov fermion, the BI vector only contributes to the anti-symmetric part. Since both parts separately have to transform covariantly under an induced diffeomorphism, this leaves only the linearly realised Poincar\'{e} symmetry. However, BI is still special amongst vector EFTs since it can form the bosonic sector of a supersymmetric theory which combines the BI vector and VA fermion. This leads to interesting multi-soft limits for BI at tree-level \cite{Cheung-vectors}.

Rather than adding higher-order inessentials, one can alternatively try to extend each of the first-order algebras above on the other side of the sequence with a central extension $C$ of the form $[P_\mu, Q_\nu] = \eta_{\mu \nu} C$. This would imply that we can solve for the vector in terms of a new scalar $\varphi$ associated to $C$ via an inverse Higgs constraint $A_\mu = \partial_\mu \varphi$; the new scalar is now the essential Goldstone. The Jacobi identitites allow for this when the first-order inessential Goldstone is a scalar or symmetric, traceless tensor. Therefore the EFTs of these algebras can be consistently truncated to their longitudinal mode, and in the symmetric, traceless case, this coincides with the special Galileon of \cite{specialgalgroup1}.

Amongst the different possibilities of section II, the algebra with the inessential two-form appears particularly interesting. Firstly, it does not allow for a scalar central extension and hence cannot be truncated to its longitudinal mode. Secondly, it is the only algebra whose degenerate limit goes beyond gauge symmetries. Finally, its non-degenerate version can be seen to be equal to a double copy of Poincar\'{e}. 
The latter suggests tantalising relations with double field theory \cite{DFT}, building on earlier results indicating factorisation of both the scattering amplitudes of gravity \cite{KLT, BCJ} and its low-energy Lagrangian \cite{Bern, Hohm, Cheung}. 

Double field theory aims to incorporate key string properties by introducing a double geometry spanned by the coordinates $x^\mu$ and their duals $\tilde x_\mu$, corresponding respectively to momentum and string winding modes. This double geometry can be seen as the generalisation of $D=5$ Minkowski, with an additional scalar coordinate (from the 4D perspective), and $D=4$ superspace, with an additional fermion coordinate. Placing a space-time filling brane along the $x^\mu$-coordinates  of double geometry would lead to the identification of $\tilde x_\mu = A_\mu$ as a vector in the worldvolume theory, and would non-linearly realise the twofold \Poincare isometries of the flat double geometry. This seems to indicate a relation to the doublet $\vec{V}^\mu$ of the symmetry algebras of section II which we leave for future investigations. 

\begin{acknowledgments}
We would like to thank Brando Bellazzini, Tomas Brauner, Sebastian Garcia-Saenz, Scott Melville, Johannes Noller, Jaroslav Trnka and Pelle Werkman for useful discussions. EM would like to thank the VSI for hospitality while this project was being completed. RK, DR and DS acknowledge the Dutch funding agency ‘Netherlands Organisation for Scientific Research’ (NWO) for financial support. EM is supported by the ERC Advanced Grant ``Strings and Gravity" (Grant No. 320045). DR is grateful for hospitality and financial support from CERN during the final stages of this research.
\end{acknowledgments}

\end{document}